\documentclass[conference]{IEEEtran}
\IEEEoverridecommandlockouts
\usepackage{cite}
\usepackage{amsmath,amssymb,amsfonts,subcaption}
\usepackage{algorithmic}
\usepackage{graphicx}
\usepackage{textcomp}
\usepackage{xcolor}
\def\BibTeX{{\rm B\kern-.05em{\sc i\kern-.025em b}\kern-.08em
    T\kern-.1667em\lower.7ex\hbox{E}\kern-.125emX}}
\begin{document}

\title{Bitrate Ladder Construction using Visual Information Fidelity}

\author{
\IEEEauthorblockN{Krishna Srikar Durbha\thanks{This research was sponsored by a grant from Meta Video Infrastructure, and by grant number 2019844 for the National Science Foundation AI Institute for Foundations of Machine Learning (IFML).}}
\IEEEauthorblockA{\textit{The University of Texas at Austin}}
\and
\IEEEauthorblockN{Hassene Tmar}
\IEEEauthorblockA{\textit{Meta Platforms, Inc.}}
\and
\IEEEauthorblockN{Cosmin Stejerean}
\IEEEauthorblockA{\textit{Meta Platforms, Inc.}}
\and
\IEEEauthorblockN{Ioannis Katsavounidis}
\IEEEauthorblockA{\textit{Meta Platforms, Inc.}}
\and
\IEEEauthorblockN{Alan C. Bovik}
\IEEEauthorblockA{\textit{The University of Texas at Austin}}
}

\maketitle

\begin{abstract}
    Recently proposed perceptually optimized per-title video encoding methods provide better BD-rate savings than fixed bitrate-ladder approaches that have been employed in the past. However, a disadvantage of per-title encoding is that it requires significant time and energy to compute bitrate ladders. Over the past few years, a variety of methods have been proposed to construct optimal bitrate ladders including using low-level features to predict cross-over bitrates, optimal resolutions for each bitrate, predicting visual quality, etc. Here, we deploy features drawn from Visual Information Fidelity (VIF) (VIF features) extracted from uncompressed videos to predict the visual quality (VMAF) of compressed videos. We present multiple VIF feature sets extracted from different scales and subbands of a video to tackle the problem of bitrate ladder construction. Comparisons are made against a fixed bitrate ladder and a bitrate ladder obtained from exhaustive encoding using Bjontegaard delta metrics.
\end{abstract}


\section{Introduction}
\label{sec:introduction}
The demand and applications for Video on Demand (VoD) have tremendously increased with the rise of video streaming services. Today video content represents more than 70 percent of global internet traffic. Videos stored on streaming platform servers are compressed and delivered to users based on their display settings, network conditions, device capabilities, etc. Over the past few years, \textit{HTTP Adaptive Streaming} has been the standard for delivering video content, and the \textit{HTTP Live Streaming} (HLS) bitrate ladder \cite{fixed-bitrate-ladder} has become the conventional method of adaptive bitrate streaming for all kinds of video content. HLS \cite{fixed-bitrate-ladder} was designed considering the vast diversity of video characteristics and network conditions. Even though `one-size-fits-all' is optimized for important video characteristics, it does not guarantee an optimal bitrate ladder for a given video.

The recently introduced per-title encoding \cite{Per-Title-Encoding, Shot-Encoding, Dynamic-Optimizer} approach provides optimizations to the existing fixed bitrate ladder (HLS) by constructing a content-optimized bitrate ladder that delivers better overall \textit{Quality of Experience} (QoE). Per-title encoding schemes design a convex hull tailored for each video, containing optimal bitrate-resolution pairs that display the highest visual quality at the target bitrate. The convex hull is where the encoding point achieves Pareto efficiency. However, the main disadvantage of this approach is the requirement for significant computation resources and time. For a set of $R$ resolutions and $B$ bitrates, constructing a Pareto-front (often referred to as convex hull in the adaptive streaming literature) requires compressing the video $R \times B$ times, then selecting the best resolution for each target bitrate. Hence, constructing the Pareto front is repeated many times to transmit a single diverse video with multiple scenes/shots.
 
As a way of reducing this complexity, while still perceptually optimizing results, we have designed an efficient approach to bitrate ladder construction using features drawn from the popular Visual Information Fidelity (VIF) \cite{VIF} video quality assessment (VQA) model. These VIF features are extracted from uncompressed videos. We concatenate these VIF features and metadata of the compressed video including bitrate, height, and width, to predict visual quality. We present multiple approaches that use VIF features extracted from different scales and subbands of videos. We evaluate the performance of these quality prediction approaches against the standard fixed bitrate ladder for HLS Authoring Specification for Apple devices \cite{fixed-bitrate-ladder} and the reference bitrate ladder obtained from the convex hull or exhaustive encoding.

\section{Related Works}
\label{sec:related-works}
To eliminate the need for exhaustive encoding in building the Pareto-front, numerous methods have been proposed to assist the construction of content-adaptive perceptually optimal bitrate ladders. In the works \cite{Predicting-Video-Rate-Distortion-Curves-using-Textural-Features},\cite{Study-of-compression-statistics-and-prediction-of-rate-distortion-curves-for-video-texture}, the authors have modeled rate-distortion (RD) curves as polynomials and used video texture features like gray-level co-occurrence matrix (GLCM), normalized cross-correlation (NCC), temporal coherence (TC), etc to predict coefficients of polynomials using Support Vector Regression models, thus demonstrating a correlation between video texture characteristics and rate-distortion curves. The same authors extended their work in \cite{Content-gnostic-Bitrate-Ladder-Prediction-for-Adaptive-Video-Streaming}, \cite{Efficient-Bitrate-Ladder-Construction-for-Content-Optimized-Adaptive-Video-Streaming}, \cite{VMAF-based-Bitrate-Ladder-Estimation-for-Adaptive-Streaming} by modeling various key points on the Pareto-front, using a similar set of video texture features to predict them. In \cite{Content-gnostic-Bitrate-Ladder-Prediction-for-Adaptive-Video-Streaming} and \cite{Efficient-Bitrate-Ladder-Construction-for-Content-Optimized-Adaptive-Video-Streaming} the authors predicted cross-over QPs between rate-quality (RQ) curves of two resolutions and used them for bitrate ladder construction. In \cite{VMAF-based-Bitrate-Ladder-Estimation-for-Adaptive-Streaming}, knee-points are predicted to identify where RQ curves have the highest curvature and these knee-points are used for the construction of bitrate ladders. Similar to these ideas, the authors in \cite{Benchmarking-Learning-based-Bitrate-Ladder-Prediction-Methods-for-Adaptive-Video-Streaming} have modeled cross-over points as bitrates and experimented using various low-level features and deep-learning feature representations to achieve good BD-rate savings against fixed and reference bitrate ladders.

In the paper \cite{Perceptually-Aware-Per-Title-Encoding-for-Adaptive-Video-Streaming}, Vignesh \textit{et al.} modeled VMAF as a linear regression depending on DCT-based energy features and bitrate, and have shown high correlation against VMAF. There have also been works \cite{Ensemble-Learning-for-Efficient-VVC-Bitrate-Ladder-Prediction}, \cite{Just-Noticeable-Difference-aware-Per-Scene-Bitrate-laddering-for-Adaptive-Video-Streaming} that directly predict optimal resolutions and CRF values using low-level video texture features and DCT-based energy features. The authors in \cite{Efficient-Per-Shot-Convex-Hull-Prediction-By-Recurrent-Learning} used deep-learning models to predict points on the Pareto-front by modeling the problem as multi-label classification. Unlike previously mentioned approaches that require extra computation for computing features used by these algorithms, our proposed approach utilizes a full-reference quality estimation (FR) model in the video-delivery pipeline to construct perceptually optimal bitrate ladders.

\section{Approach}
\label{sec:pagestyle}
Visual Multimethod Assessment Fusion (VMAF) \cite{VMAF} is a widely used full reference VQA model that achieves excellent correlations against human judgments. VMAF produces quality scores by fusing features from VIF \cite{VIF}, the average absolute luminance differences between adjacent frames, and the Detail Loss Metric (DLM) \cite{DLM}. VIF quantifies the information that could ideally be extracted by the brain from the reference image and determines the loss of information from distortion. It uses a Gaussian scale mixture image model expressed in the wavelet domain. We utilize spatial VIF features obtained from frame-based VIF, the mean absolute luminance differences computed by VMAF, and the same VIF features computed on the full frame differences. Hence, these feature sets can be calculated using the existing VMAF pipeline. The following equations explain the Gaussian Scale Mixture model of a subband of the reference image:
\begin{align}
    \mathcal{C} = \mathcal{S} . \mathcal{U} = \{S_{i}.\vec{U}_{i}: i \in I\}\\
    \mathcal{E} = \mathcal{C} + \mathcal{N} \\
    I(\vec{C}^N;\vec{E}^N | s^N) = \sum_{j=1}^{N}\sum_{i=1}^{N} I(\vec{C}_{i};\vec{E}_{j} | \vec{C}^{i-1}, \vec{E}^{j-1}, s^N) \\
    I(\vec{C}^N;\vec{E}^N | S^N = s^N) = \sum_{i=1}^{N} I(\vec{C}_{i};\vec{E}_{i} | s_{i}) \\
    I(\vec{C}^N;\vec{E}^N | s^N) = \frac{1}{2}\sum_{i=1}^{N} \log_{2}(\frac{|s_{i}^2\mathbf{C_{U}} + \sigma_{n}^{2}\mathbf{I}|}{|\sigma_{n}^{2}\mathbf{I}|}), \label{eqn:vif0}
\end{align}
where $\mathcal{C} = \{\vec{C}_{i}: i \in I\}$ is a random field (RF) representing a subband of the reference image, $\mathcal{E} = \{\vec{E}_{i}: i \in I\}$ is the output of the HVS model, $\mathcal{S} = \{\vec{S}_{i}: i \in I\}$ is a RF of positive scalars, $\mathcal{U} = \{\vec{U}_{i}: i \in I\}$ is a Gaussian RF with mean zero and covariance $\mathbf{C}_{U}$ and $\mathcal{N} = \{\vec{N}_{i}: i \in I\}$ is HVS noise in the wavelet domain and is a multivariate Gaussian with mean zero and covariance $\mathbf{C}_{N} = \sigma_{n}^2\mathbf{I}$. $\vec{C}_{i}$ and $\vec{U}_{i}$ are M-dimensional vectors and $\vec{U}_{i}$ is independent of $\vec{U}_{j}$ for $i \neq j$. $\vec{s}^N$ denotes the realization of $S^N = (S_{1}, S_{2}, \dots, S_{N})$ for a particular reference image, which can be thought of as model parameters associated with it. On matrix factorization of \eqref{eqn:vif0}, we get:
\begin{align}
    I(\vec{C}^N;\vec{E}^N | s^N) = \frac{1}{2}\sum_{i=1}^{N}\sum_{j=1}^{M} \log_{2}(1 + \frac{s_{i}^2\lambda_{j}}{\sigma_{n}^{2}}).
\end{align}
$I(\vec{C}^N;\vec{E}^N | s^N)$ represents the information that could ideally be extracted by the brain from a particular subband in the image. We calculate all the above features over four scales, each having two subbands, using $M = 9$. We consider $I_{k,b}^{j}$ to represent mutual information along the $j^{th}$ eigenvector of the $b^{th}$ subband for the $k^{th}$ scale, where $j \in \{1,2,...,M\}$ , $b \in \{1,2\}$ and $k \in \{1,2,3,4\}$:
\begin{align}
    I_{k,b}^{j} = \frac{1}{N}\sum_{i=1}^{N} \log_{2}(1 + \frac{s_{i}^2\lambda_{j}}{\sigma_{n}^{2}}) \\
    I_{k,b} = \frac{1}{N}\sum_{j=1}^{M}\sum_{i=1}^{N} \log_{2}(1 + \frac{s_{i}^2\lambda_{j}}{\sigma_{n}^{2}}) \\
    I_{k} = \frac{1}{2}\sum_{b=1}^{2}I_{k,b}.
\end{align}

\begin{table}
    \renewcommand{\arraystretch}{1.5}
    \centering
    \begin{tabular}{| c | c | c | } 
      \hline
      \textbf{Approach} & \textbf{Features Set} & \textbf{No.of Features} \\ 
      \hline
      1 & $I_{k}[F_{i}]$, $b$, $\frac{w}{3840}$, $\frac{h}{3840}$ & 7 \\ 
      \hline
      2 & $I_{k,b}[F_{i}]$, $b$, $\frac{w}{3840}$, $\frac{h}{3840}$ & 11\\ 
      \hline
      3 & $I_{k,b}^{j}[F_{i}]$, $b$, $\frac{w}{3840}$, $\frac{h}{3840}$ & 75\\ 
      \hline
      4 & $I_{k}[F_{i}]$, $|D_{i}|$, $b$, $\frac{w}{3840}$, $\frac{h}{3840}$ & 8 \\ 
      \hline
      5 & $I_{k,b}[F_{i}]$, $|D_{i}|$, $b$, $\frac{w}{3840}$, $\frac{h}{3840}$ & 12\\ 
      \hline
      6 & $I_{k,b}^{j}[F_{i}]$, $|D_{i}|$, $b$, $\frac{w}{3840}$, $\frac{h}{3840}$ & 76\\ 
      \hline
      7 & $I_{k}[F_{i}]$, $|D_{i}|$, $I_{k}[D_{i}]$, $b$, $\frac{w}{3840}$, $\frac{h}{3840}$ & 12 \\ 
      \hline
      8 & $I_{k,b}[F_{i}]$, $|D_{i}|$, $I_{k,b}[D_{i}]$, $b$, $\frac{w}{3840}$, $\frac{h}{3840}$ & 20\\ 
      \hline
      9 & $I_{k,b}^{j}[F_{i}]$, $|D_{i}|$, $I_{k,b}^{j}[D_{i}]$, $b$, $\frac{w}{3840}$, $\frac{h}{3840}$ & 148\\ 
      \hline
    \end{tabular}
    \caption{VIF feature sets}
    \label{table:VIF-Features}
\end{table}

Since VIF only captures spatial characteristics, we also compute the mean absolute luminance differences between frames, as used in VMAF, and the same VIF features as explained above, but differences between consecutive frames. Table \ref{table:VIF-Features} shows the nine different feature sets used in our experiments, where $D_{i} = F_{i} - F_{i-1}$. We temporally pool each of the VIF-based features by calculating their mean, then concatenate them with the following metadata: bitrate $b$, scaled width $w/3840$, and height $h/3840$. All these measurements are made on each compressed video whose visual quality needs to be predicted. 

\section{Evaluation and Results}
\label{sec:evaluation}
\begin{table*}[!ht]
    \renewcommand{\arraystretch}{1.2}
    \centering
    \caption{Table containing the mean and standard deviation values of BD-rate and BD-VMAF showing the performance of the Extra-Trees regressor against fixed and reference bitrate ladders on all the videos in validation and test datasets.}
    \begin{tabular}{ c | c  c | c  c}
    \hline
    Features Set & \multicolumn{2}{c}{BL vs Fixed Bitrate Ladder} & \multicolumn{2}{c}{BL vs Reference Bitrate Ladder}\\
    \hline
     & BD-Rate (in \%) & BD-VMAF & BD-Rate (in \%) & BD-VMAF\\
    \hline
    $I_{k}[F_{i}]$, $b$, $\frac{w}{3840}$, $\frac{h}{3840}$ & $-15.173/14.823$ & $3.662/3.737$ & $3.318/4.886$ & $-0.693/0.948$\\
    
    $I_{k,b}[F_{i}]$, $b$, $\frac{w}{3840}$, $\frac{h}{3840}$ &  $-15.097/14.214$ & $3.694/3.65$ & $3.404/4.598$ & $-0.673/0.852$\\
    
    $I_{k,b}^{j}[F_{i}]$, $b$, $\frac{w}{3840}$, $\frac{h}{3840}$ &  $\mathbf{-15.188}/13.6$ & $3.705/3.557$ & $3.425/4.484$ & $-0.676/0.764$ \\
    
    $I_{k}[F_{i}]$, $|D_{i}|$, $b$, $\frac{w}{3840}$, $\frac{h}{3840}$ &  $-14.73/15.181$ & $3.625/3.876$ & $3.763/5.741$ & $-0.753/1.101$\\
    
    $I_{k,b}[F_{i}]$, $|D_{i}|$, $b$, $\frac{w}{3840}$, $\frac{h}{3840}$ &  $-15.03/14.652$ & $3.67/3.741$ & $3.485/4.942$ & $-0.701/0.947$\\
    
    $I_{k,b}^{j}[F_{i}]$, $|D_{i}|$, $b$, $\frac{w}{3840}$, $\frac{h}{3840}$ &  $-14.756/14.26$ & $3.625/3.673$ & $3.91/5.055$ & $-0.757/0.86$\\
    
    $I_{k}[F_{i}]$, $|D_{i}|$, $I_{k}[D_{i}]$, $b$, $\frac{w}{3840}$, $\frac{h}{3840}$ &  $-14.355/15.87$ & $\mathbf{3.783}/3.84$ & $3.735/5.443$ & $\mathbf{-0.602}/0.677$\\
    
$I_{k,b}[F_{i}]$, $|D_{i}|$, $I_{k,b}[D_{i}]$, $b$, $\frac{w}{3840}$, $\frac{h}{3840}$ &  $-15.187/14.664$ & $3.731/3.821$ & $\mathbf{3.307}/4.644$ & $-0.664/0.84$\\
    
    $I_{k,b}^{j}[F_{i}]$, $|D_{i}|$, $I_{k,b}^{j}[D_{i}]$, $b$, $\frac{w}{3840}$, $\frac{h}{3840}$ &  $-15.071/14.202$ & $3.672/3.805$ & $3.441/4.026$ & $-0.715/0.694$\\
    \hline
    \end{tabular}
    \label{table: bd-metrics}
\end{table*}

\subsection{Dataset Preparation}
We used the BVT-100 4K dataset used in \cite{Content-gnostic-Bitrate-Ladder-Prediction-for-Adaptive-Video-Streaming}, \cite{Efficient-Bitrate-Ladder-Construction-for-Content-Optimized-Adaptive-Video-Streaming} in our experiments. The video sequences were drawn from various public sources including Netflix Chimera, Ultra Video Group, Harmonic Inc., SJTU, and AWS Elemental. All the sequences were spatially cropped to UHD (3840 x 2160 pixels), converted to 4:2:0 chroma subsampling, and temporally cropped to 64 frames. Each sequence is drawn from a single scene (without scene cuts) and the majority of the test sequences have a frame rate of 60 fps and a bit depth of 10 bits per sample. We divided the dataset into three non overlapping sets for training (70 videos), validation (10 videos), and testing (20 videos). We ensured that there was no match in video titles between each of these sets to create maximum diversity of video statistics. We used \textit{ffmpeg} to conduct compression and quality estimation. The videos were compressed using the \textbf{libx265} codec with the \textbf{medium} preset using resolutions $\{3840 \times 2160, 2560 \times 1440, 1920 \times 1080, 1280 \times 720, 960 \times 540, 768 \times 432, 640 \times 360, 512 \times 288\}$ and performed constant-quality encoding by varying the CRF values from 18 to 50 (inclusive) on each resolution. We computed full-reference quality metrics like VIF and VMAF after the compressed video was upscaled to the original resolution ($3840 \times 2160$). We used the Lanczos algorithm for upscaling and downscaling (from 2160p) of videos.

\subsection{Training and Prediction}
In all the experiments, we considered bitrates on a log scale with base $2$ and scaled VMAF by 100 so that it lies in [0,1]. We trained various versions of each model using multiple regressors including Extra-Trees, XG-Boost, and Random-Forests. Machine learning models are trained separately for each feature set/approach, and similar to \cite{Benchmarking-Learning-based-Bitrate-Ladder-Prediction-Methods-for-Adaptive-Video-Streaming}, we observed that the Extra-Trees regressor performed the best in all our feature sets consistently.

\subsection{Evaluation}
We evaluated the performance of each approach on the validation and test datasets for a bigger sample size. For evaluation, we constructed a bitrate ladder for bitrates (in Mbps) $\{0.25, 0.5, 1, 2, 3, 4, 5, 6, 7, 8, 9, 10.5\}$. Based on predictions produced by the quality prediction models, we selected the optimal resolution for a bitrate as the one that yielded the highest-quality prediction. Since machine learning models are associated with an error during prediction, and we need our bitrate ladder to be monotonic, we applied a simple correction to our predicted bitrate ladder. The algorithm traverses the bitrate ladder from higher bitrates to low bitrates and imposes a condition that the optimal resolution at the current bitrate would be less than or equal to the optimal resolution observed at the previous (next higher) bitrate. Since we are using constant-quality encoding, if $b_{i}$ and $R_{i}$ are steps on the bitrate ladder, for all rate-quality (RQ) points encoded at $R_{i}$, we selected the RQ point closest to $b_{i}$ to be a point on the convex hull. We compared the Bjontegaard delta metrics of our proposed approaches against a fixed bitrate ladder \cite{fixed-bitrate-ladder} and a reference bitrate ladder we constructed by exhaustive encoding.

\begin{figure}[!ht]
    \centering
    \includegraphics[width=0.975\columnwidth]{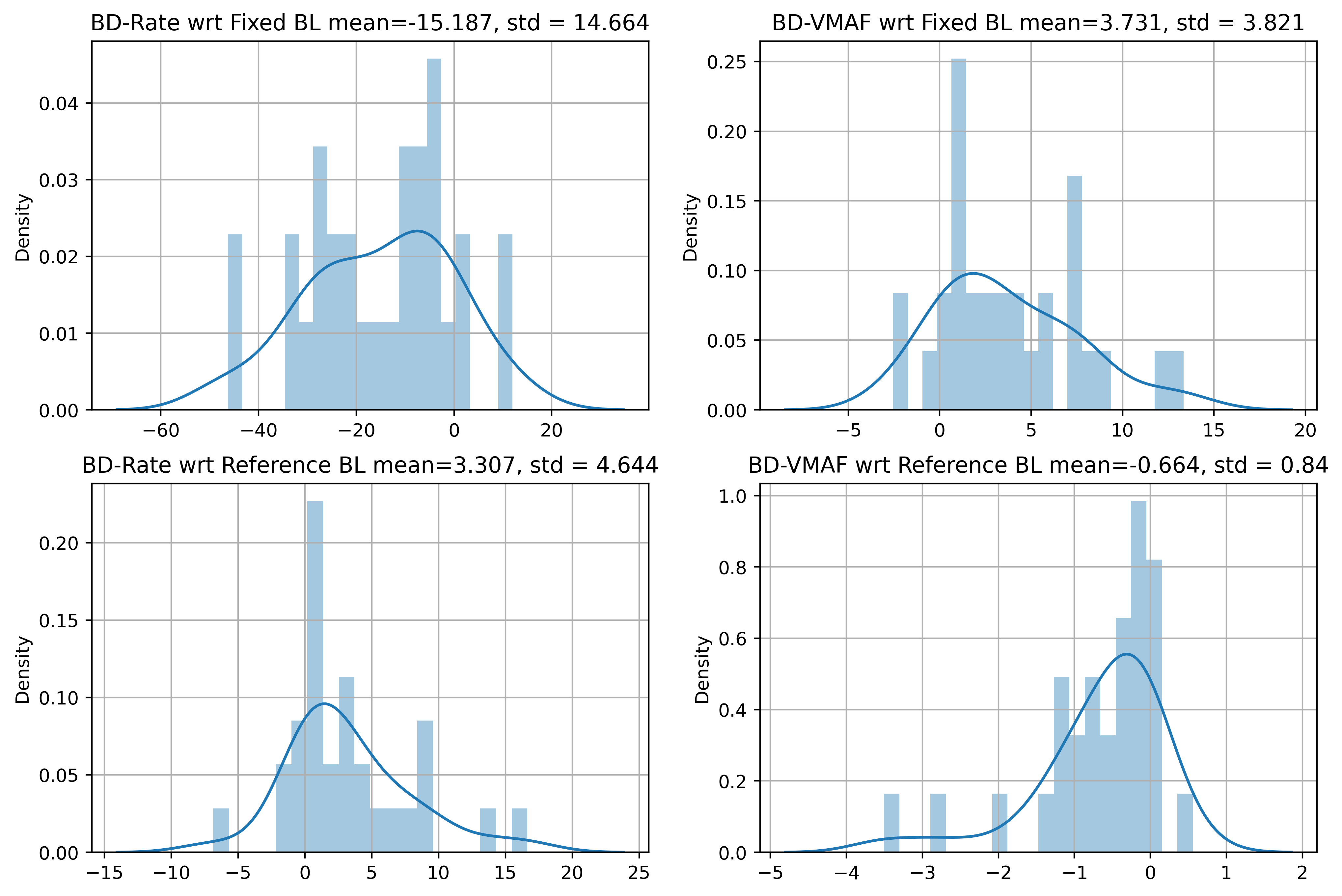}
    \caption{Distributions of BD-rate and BD-VMAF for bitrate ladder constructed with Approach-8 against fixed and reference bitrate for all videos in validation and test datasets.}
    \label{fig: Approach-8 Histogram}
\end{figure}

\begin{figure}[!ht]
    \centering
    \begin{subfigure}[b]{0.77\columnwidth}
        \includegraphics[width=\linewidth]{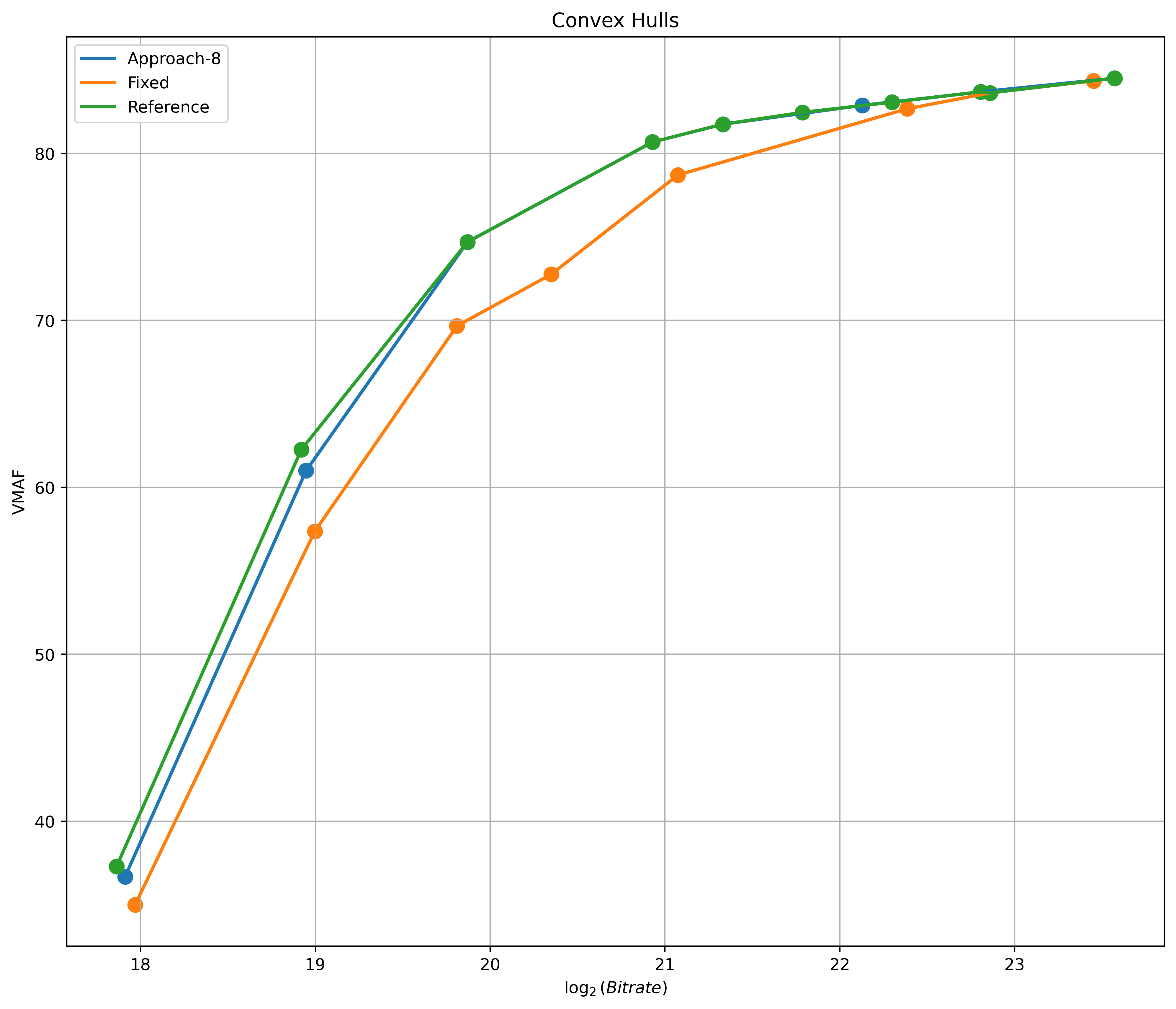}
        \caption{Honeybee}
    \end{subfigure}
    \hfill
    \begin{subfigure}[b]{0.77\columnwidth}
        \includegraphics[width=\linewidth]{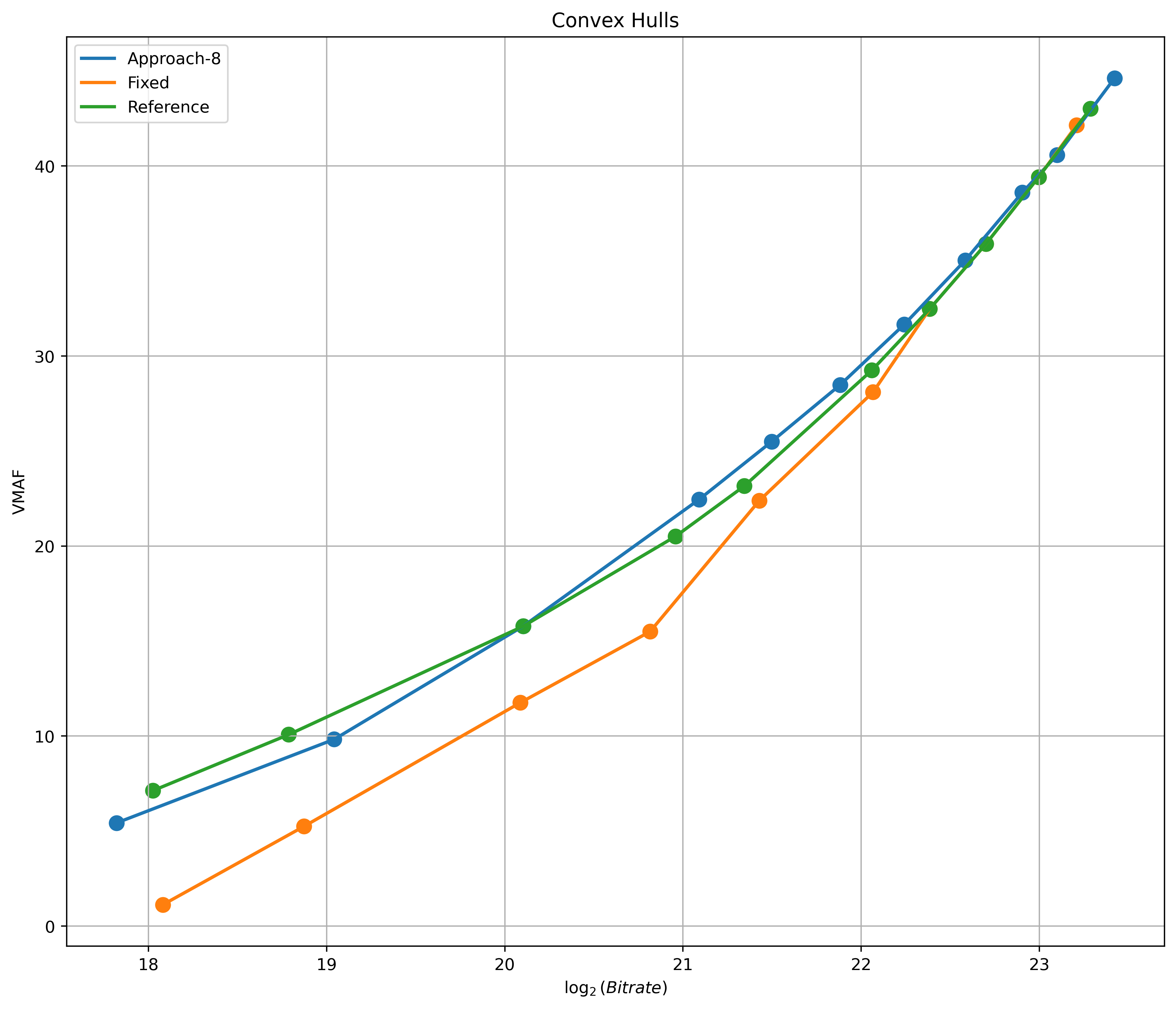}
        \caption{Fountains}
    \end{subfigure}
    \begin{subfigure}[b]{0.77\columnwidth}
        \includegraphics[width=\linewidth]{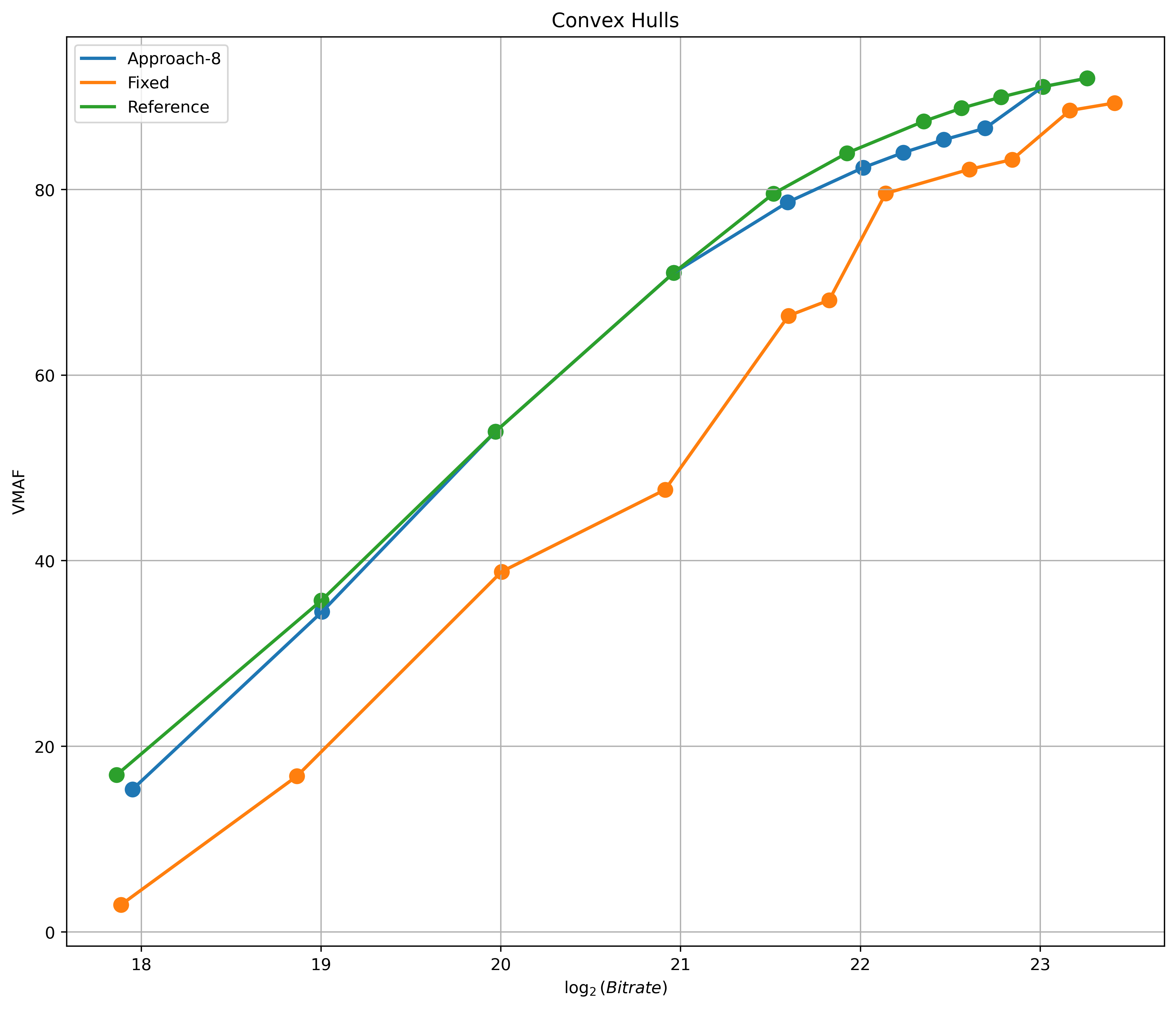}
        \caption{Tallbuildings}
  \end{subfigure}
  \caption{Convex Hulls constructed using Approach-8, fixed bitrate ladder, and reference bitrate ladder for three different video files from the test dataset.}
  \label{fig:convex-hulls}
\end{figure}

The reference bitrate ladder has a mean BD-rate $-17.95\%$ (negative BD-rate means a gain in bitrate) with standard-deviation of $13.058\%$, and a mean BD-VMAF of 4.38 (positive means gain in quality) with standard-deviation of 3.575 against the fixed bitrate ladder. Table \ref{table: bd-metrics} shows the performances of the compared sets of features using BD-rate and BD-VMAF values. All the approaches demonstrate significant gains against the fixed bitrate ladder and performed very close to the reference bitrate ladder. The Extra-Trees regressor trained on features $I_{k,b}[F_{i}]$, $|D_{i}|$, $I_{k,b}[D_{i}]$, $b$, $\frac{w}{3840}$, $\frac{h}{3840}$ (Approach-8) obtained the best performance considering all BD-metrics. Fig.\ref{fig: Approach-8 Histogram} plots the histograms of the BD-metrics and Fig.\ref{fig:convex-hulls} plots convex hulls constructed using Approach-8, fixed and reference bitrate ladders.

\section{Conclusion}
\label{sec:conclusion}
We deployed features drawn from Visual Information Fidelity to define feature sets extracted from different scales and subbands. We used these feature sets, along with metadata describing bitrate, width, and height, to predict the visual quality of compressed videos without performing any additional compression or quality estimation. Our method utilizes the FR quality estimation block of the video-delivery pipeline for feature extraction and hence can easily be integrated into existing video delivery pipelines. We used VIF-based quality prediction models to construct an optimal content-optimized perceptual bitrate ladder. Our best-performing approach showed average BD-rate and BD-VMAF gains of 15.187\% and 3.731 respectively, against Apple's fixed bitrate ladder and BD-rate and BD-VMAF losses of 3.307\% and 0.664 respectively, against a bitrate ladder obtained by exhaustive encoding.

\bibliographystyle{IEEEtran}
\bibliography{refs}

\begin{thebibliography}{10}
\providecommand{\url}[1]{#1}
\csname url@samestyle\endcsname
\providecommand{\newblock}{\relax}
\providecommand{\bibinfo}[2]{#2}
\providecommand{\BIBentrySTDinterwordspacing}{\spaceskip=0pt\relax}
\providecommand{\BIBentryALTinterwordstretchfactor}{4}
\providecommand{\BIBentryALTinterwordspacing}{\spaceskip=\fontdimen2\font plus
\BIBentryALTinterwordstretchfactor\fontdimen3\font minus
  \fontdimen4\font\relax}
\providecommand{\BIBforeignlanguage}[2]{{%
\expandafter\ifx\csname l@#1\endcsname\relax
\typeout{** WARNING: IEEEtran.bst: No hyphenation pattern has been}%
\typeout{** loaded for the language `#1'. Using the pattern for}%
\typeout{** the default language instead.}%
\else
\language=\csname l@#1\endcsname
\fi
#2}}
\providecommand{\BIBdecl}{\relax}
\BIBdecl

\bibitem{fixed-bitrate-ladder}
\BIBentryALTinterwordspacing
``Http live streaming (hls) authoring specification for apple devices.''
  [Online]. Available:
  \url{https://developer.apple.com/documentation/http-live-streaming/hls-authoring-specification-for-apple-devices}
\BIBentrySTDinterwordspacing

\bibitem{Per-Title-Encoding}
\BIBentryALTinterwordspacing
``Per-title encode optimization.'' [Online]. Available:
  \url{https://netflixtechblog.com/per-title-encode-optimization-7e99442b62a2}
\BIBentrySTDinterwordspacing

\bibitem{Shot-Encoding}
\BIBentryALTinterwordspacing
``Optimized shot-based encodes.'' [Online]. Available:
  \url{https://netflixtechblog.com/optimized-shot-based-encodes-now-streaming-4b9464204830}
\BIBentrySTDinterwordspacing

\bibitem{Dynamic-Optimizer}
\BIBentryALTinterwordspacing
``Dynamic optimizer.'' [Online]. Available:
  \url{https://netflixtechblog.com/dynamic-optimizer-a-perceptual-video-encoding-optimization-framework-e19f1e3a277f}
\BIBentrySTDinterwordspacing

\bibitem{VIF}
H.~Sheikh and A.~Bovik, ``Image information and visual quality,'' \emph{IEEE
  Transactions on Image Processing}, vol.~15, no.~2, pp. 430--444, 2006.

\bibitem{Predicting-Video-Rate-Distortion-Curves-using-Textural-Features}
A.~V. Katsenou, M.~Afonso, D.~Agrafiotis, and D.~R. Bull, ``Predicting video
  rate-distortion curves using textural features,'' in \emph{2016 Picture
  Coding Symposium, {PCS} 2016, Nuremberg, Germany, December 4-7, 2016}.

\bibitem{Study-of-compression-statistics-and-prediction-of-rate-distortion-curves-for-video-texture}
A.~V. Katsenou, M.~Afonso, and D.~R. Bull, ``Study of compression statistics
  and prediction of rate-distortion curves for video texture.''

\bibitem{Content-gnostic-Bitrate-Ladder-Prediction-for-Adaptive-Video-Streaming}
A.~V. Katsenou, J.~Sole, and D.~R. Bull, ``Content-gnostic bitrate ladder
  prediction for adaptive video streaming,'' in \emph{Picture Coding Symposium,
  {PCS} 2019, Ningbo, China, November 12-15, 2019}.

\bibitem{Efficient-Bitrate-Ladder-Construction-for-Content-Optimized-Adaptive-Video-Streaming}
A.~V. Katsenou, J.~Sole, and D.~Bull, ``Efficient bitrate ladder construction
  for content-optimized adaptive video streaming,'' \emph{IEEE Open Journal of
  Signal Processing}, vol.~2, pp. 496--511, 2021.

\bibitem{VMAF-based-Bitrate-Ladder-Estimation-for-Adaptive-Streaming}
A.~V. Katsenou, F.~Zhang, K.~Swanson, M.~Afonso, J.~Sole, and D.~R. Bull,
  ``Vmaf-based bitrate ladder estimation for adaptive streaming,'' in
  \emph{Picture Coding Symposium, {PCS} 2021, Bristol, United Kingdom, June 29
  - July 2, 2021}.

\bibitem{Benchmarking-Learning-based-Bitrate-Ladder-Prediction-Methods-for-Adaptive-Video-Streaming}
A.~Telili, W.~Hamidouche, S.~A. Fezza, and L.~Morin, ``Benchmarking
  learning-based bitrate ladder prediction methods for adaptive video
  streaming,'' in \emph{Picture Coding Symposium, {PCS} 2022, San Jose, CA,
  USA, December 7-9, 2022}.

\bibitem{Perceptually-Aware-Per-Title-Encoding-for-Adaptive-Video-Streaming}
V.~V. Menon, H.~Amirpour, M.~Ghanbari, and C.~Timmerer, ``Perceptually-aware
  per-title encoding for adaptive video streaming,'' in \emph{{IEEE}
  International Conference on Multimedia and Expo, {ICME} 2022, Taipei, Taiwan,
  July 18-22, 2022}.

\bibitem{Ensemble-Learning-for-Efficient-VVC-Bitrate-Ladder-Prediction}
F.~Nasiri, W.~Hamidouche, L.~Morin, N.~Dhollande, and J.~Aubi{\'{e}},
  ``Ensemble learning for efficient {VVC} bitrate ladder prediction,'' in
  \emph{10th European Workshop on Visual Information Processing, Lisbon,
  Portugal, September 11-14, 2022}.

\bibitem{Just-Noticeable-Difference-aware-Per-Scene-Bitrate-laddering-for-Adaptive-Video-Streaming}
V.~V. Menon, J.~Zhu, P.~T. Rajendran, H.~Amirpour, P.~L. Callet, and
  C.~Timmerer, ``Just noticeable difference-aware per-scene bitrate-laddering
  for adaptive video streaming,'' \emph{CoRR}, vol. abs/2305.00225, 2023.

\bibitem{Efficient-Per-Shot-Convex-Hull-Prediction-By-Recurrent-Learning}
S.~Paul, A.~Norkin, and A.~C. Bovik, ``Efficient per-shot convex hull
  prediction by recurrent learning,'' \emph{CoRR}, vol. abs/2206.04877, 2022.

\bibitem{VMAF}
\BIBentryALTinterwordspacing
``Vmaf - video multi-method assessment fusion.'' [Online]. Available:
  \url{https://github.com/Netflix/vmaf}
\BIBentrySTDinterwordspacing

\bibitem{DLM}
S.~Li, F.~Zhang, L.~Ma, and K.~N. Ngan, ``Image quality assessment by
  separately evaluating detail losses and additive impairments,'' \emph{IEEE
  Transactions on Multimedia}, vol.~13, no.~5, pp. 935--949, 2011.

\end{thebibliography}

\end{document}